\documentclass[prc,aps,superscriptaddress,twocolumn,showpacs,nofootinbib]{revtex4}

\usepackage{graphicx,amsmath,bm,multirow}
\usepackage{amsthm}
\usepackage{amsfonts,mathrsfs,amssymb}

\usepackage[usenames,dvipsnames]{color}
\usepackage[colorlinks,linkcolor=blue,anchorcolor=blue,citecolor=blue]{hyperref}

\newcommand{\br}{{\mathbf{r}}}
\newcommand{\bq}{{\mathbf{q}}}

\newcommand{\bx}{{\mathbf{x}}}

\newcommand{\beq}{\begin{equation}}
\newcommand{\eeq}{\end{equation}}
\newcommand{\beqn}{\begin{eqnarray}}
\newcommand{\eeqn}{\end{eqnarray}}
\newcommand{\bsub}{\begin{subequations}}
\newcommand{\esub}{\end{subequations}}
\newcommand{\bpm}{\begin{pmatrix}}
\newcommand{\epm}{\end{pmatrix}}

\newcommand{\ti}{\text{i}}

\newcommand{\ring}[1]{\textcolor{black}{#1}}

\begin{document}
\title{Systematic study of nuclear matrix elements in neutrinoless double-beta decay with a beyond mean-field covariant density functional theory}

\author{J. M. Yao}
 \affiliation{Department of Physics, Tohoku University, Sendai 980-8578, Japan}
 \affiliation{School of Physical Science and Technology, Southwest University, Chongqing 400715, China}
 \affiliation{Kavli Institute for Theoretical Physics China, Chinese Academy of Sciences, Beijing 100190, China}

\author{L. S. Song}
 \affiliation{State Key Laboratory of Nuclear Physics and Technology, School of Physics, Peking University, Beijing 100871, China}

\author{K. Hagino}
 \affiliation{Department of Physics, Tohoku University, Sendai 980-8578, Japan}

\author{P. Ring}
 \affiliation{Physik Department, Technische Universit\"{a}t M\"{u}nchen, D-85747 Garching, Germany}
 \affiliation{State Key Laboratory of Nuclear Physics and Technology, School of Physics, Peking University, Beijing 100871, China}
 \affiliation{Kavli Institute for Theoretical Physics China, Chinese Academy of Sciences, Beijing 100190, China}

\author{J. Meng}
 \affiliation{State Key Laboratory of Nuclear Physics and Technology, School of Physics, Peking University, Beijing 100871, China}
 \affiliation{School of Physics and Nuclear Energy Engineering, Beihang University,
              Beijing 100191, China}
 \affiliation{Department of Physics, University of Stellenbosch, Stellenbosch 7602, South Africa}

\date{\today}

%
\begin{abstract}
We report a systematic study of nuclear matrix elements (NMEs) in neutrinoless double-beta decays with a state-of-the-art beyond mean-field covariant density functional theory. The dynamic effects of particle-number and angular-momentum conservations as well as quadrupole shape fluctuations are taken into account with projections and generator coordinate method for both initial and final nuclei. The full relativistic transition operator is adopted to calculate the NMEs. The present systematic studies show that in most of the cases there is a much better agreement with the previous non-relativistic calculation based on the Gogny force than in the case of the nucleus $^{150}$Nd found in Song et al. [Phys. Rev. C 90, 054309 (2014)]. In particular, we find that the total NMEs can be well approximated by the pure axial-vector coupling term with a considerable reduction of the computational effort.
\end{abstract}
%
%


\maketitle

%
%

 \section{Introduction}
 The neutrinoless double beta ($0\nu\beta\beta$ ) decay is a process where an even-even nucleus decays into the even-even neighbor with two neutrons less and two protons more emitting only two electrons. The search of this lepton-number-violating (LNV) process in atomic nuclei is one of the current main experimental goals in nuclear and particle physics.
This LNV process occurs only if the neutrino is a Majorana particle. In particular, several fundamental questions about the nature of a neutrino, such as its absolute mass scale and mass hierarchy, are expected to be answered by combining the results from the measurements of this process and neutrino oscillations~\cite{Fukuda98,Ahmad02,Eguchi03,An12}. To date, the $0\nu\beta\beta$-decay has not been detected except for the controversial claim of detection in $^{76}$Ge by the Heiderlberg-Moscow collaboration~\cite{Klapdor04} that has recently been overruled by the constraints from the cosmology observations~\cite{Hannestad10,Komatsu11} and the latest data released by the EXO-200, KamLAND-Zen and GERDA collaborations~\cite{Auger12,Gando13,Agostini13}.

According to the neutrino mass mechanism of exchange light Majorana neutrinos, the inverse of the half-life $T^{0\nu}_{1/2}$ of the $0\nu\beta\beta$ decay process is directly related to the effective Majorana neutrino mass~\cite{Haxton84,Doi85,Tomoda91}
\begin{eqnarray}
\label{half-life}
    \left[T_{1/2}^{0\nu}\right]^{-1}=G_{0\nu} g^4_A(0)
    \left|\frac{\langle m_{\beta\beta}\rangle}{m_e}\right|^2{\left|M^{0\nu}(0_I^+\rightarrow 0_F^+)\right|^2},
\end{eqnarray}
where the axial-vector coupling constant $g_A(0)$ and the electron mass $m_e$ are constants, and the kinematic phase-space factor $G_{0\nu}$ can be determined precisely~\cite{Kotila12}. An accurate value of the nuclear matrix element (NME) $M^{0\nu}$ is essential for determining the effective neutrino mass $\langle m_{\beta\beta}\rangle$ if the decay rate is eventually measured. Although the  $0\nu\beta\beta$ decay has not bee observed yet, the NME can provide a constraint on the upper limit for the effective neutrino mass based on the current data on the lower limit of $T^{0\nu}_{1/2}$. Inversely, together with the constraints on the neutrino mass from other measurements, the NME can provide a lower limit on the half-life of the $0\nu\beta\beta$-decay, which serves as a guideline for the development of ``next-generation" experiments~\cite{Barabash2011PPN}. In any case, an accurate knowledge of the NME
for the $0\nu\beta\beta$-decay is therefore very important in nuclear physics, particle physics and cosmology~\cite{Suhonen98,Faessler98,Avignone08,Bilenky12,Vergados12,Vogel12}.

The calculation of the NME requires two main ingredients. One is the wave functions of the initial and final states, which have been calculated based on different nuclear models, including configuration-interaction shell model (ISM)~\cite{Caurier2008,Menendez2009,Neacsu2012,Horoi2013}, quasi-particle random phase approximation (QRPA)~\cite{Simkovic1999,Kortelainen2007,Fang2010,Faessler12,Mustonen2013,Terasaki2014}, interacting boson model (IBM)~\cite{Barea2009}, angular momentum projected (AMP) Hartree-Fock-Bogoliubov (PHFB) theory based on a schematic Hamiltonian~\cite{Rath2010}, and the beyond mean-field density functional theory (BMF-DFT) based on a non-relativistic energy density functional (NREDF) Gogny force~\cite{Rodriguez2010,Vaquero2013}. Compared with the PHFB, the BMF-DFT includes additional correlations connected with particle number projection (PNP), as well as fluctuations in quadrupole shapes~\cite{Rodriguez2010} and pairing gaps~\cite{Vaquero2013}. Another important ingredient is the decay operator, which reflects the mechanism of decay process. All the previous calculations of the NMEs are based on non-relativistic frameworks, \ring{adopting non-relativistic reduced transition operators derived from charge-changing nuclear currents. The resulting NMEs in these model calculations differ from each other up to a factor of $2-3$. Understanding the origin of this discrepancy has become the main goal of future studies on this topic. In particular, it is not clear whether the NMEs are sensitive to the EDF adopted in the BMF-DFT study and this question should definitely be investigated.}

In recent years, we have established a beyond mean-field covariant density functional theory (BMF-CDFT), which has been successfully applied to study many interesting phenomena related to nuclear low-lying states~\cite{Yao10}. In this paper, we report a systematic calculation of nuclear structural properties and the NMEs for the popularly studied $0\nu\beta\beta$ decay candidate nuclei within this relativistic framework. The full relativistic transition operators derived from the one-body charge-changing nuclear current, together with the ground-state wave functions from the BMF-CDFT~\cite{Yao10} are adopted in the calculation of the NMEs.  Detailed formulism and a proof-of-principle calculation for the $0\nu\beta\beta$ decay in $^{150}$Nd have been described in Ref.~\cite{Song14}.

The aim of this paper is to present systematic calculations of NMEs based on a relativistic energy density functional (REDF) and to address the open question on the sensitivity of the NMEs to the underlying EDF by making a detailed comparison with the previous BMF calculations based on the NREDF of Gogny force~\cite{Rodriguez2010}. As we show in the present work, in most of the cases there is a much better agreement between NREDF and REDF calculated matrix elements than the case of $^{150}$Nd investigated in Ref.~\cite{Song14}. This nucleus is close to the phase transition X(5) and therefore very sensitive to details of the model. Its complex nuclear structure has been described differently within these two frameworks~\cite{Niksic07,Rodriguze08}.  Moreover we analyse the net contribution of the relativistic effects and tensor terms that have been neglected in the NREDF study of Ref.~\cite{Rodriguez2010}, together with the effect of particle-number conservation that was neglected in most QRPA and PHFB studies. In particular, we find that the total NMEs can be well approximated with only the axial-vector coupling term in the charge-changing nuclear current. This provides a considerable reduction of computational efforts for future studies the NMEs.

This paper is organized in the following way. In Sec.~\ref{Formalism}, we present a brief introduction to the formalism adopted to calculate the NMEs with our BMF-CDFT. Sec.~\ref{numerical} gives the numerical details in the calculations. In Sec.~\ref{result} we present the results of the systematic study on the nuclear structural properties and the NMEs. Our findings are summarized in Sec.~\ref{Summary}.

 \section{Formalism}
 \label{Formalism}
 In our BMF-CDFT, nuclear many-body wave functions of low-lying states in initial ($I$) mother  or  final  ($F$) daughter nuclei are given as linear combinations of particle number $N, Z$, and angular momentum $J$ projected relativistic mean-field (RMF) plus BCS wave functions $\vert \beta \rangle$ constrained to have different intrinsic axial deformation $\beta$,
 \begin{equation}
 \label{GCM:wf}
 \vert JMNZ; \alpha\rangle
 =\sum_\beta f^{JNZ}_\alpha(\beta)\hat P^J_{MK=0} \hat P^N\hat P^Z\vert \beta \rangle,
 \end{equation}
 where $\alpha$ is a label distinguishing the states with same quantum numbers $J, N$, and $Z$. The $\hat P^J_{MK}$ and $\hat P^{N,Z}$ are projection operators onto angular momentum $J$ and particle number of neutrons or protons, respectively. This method is also referred to as GCM+PNAMP based on CDFT or multi-reference CDFT. The weight function $f_\alpha^{JNZ}(\beta)$ is determined from the minimization of the total energy of the state, which leads to Hill-Wheeler-Griffin (HWG) equation~\cite{Ring80}. The solution of the HWG equation provides the energy spectra and all the information needed for calculating the electric multipole transition strengths in the initial or final nuclei. We note that all the observables are calculated in full model space of occupied single-particle states.

 With the ground-state wave functions $|0_{I/F}^+\rangle$ of the initial and final nuclei from the BMF-CDFT calculation, the NME $M^{0\nu}$  for the $0\nu\beta\beta$-decay can be calculated straightforwardly as follows~\cite{Avignone08,Song14}
\beqn
\label{NME:formula}
  M^{0\nu}
  &=&\dfrac{4\pi R}{g^2_A(0)}\int\int d^3x_1\int d^3x_2 \int \dfrac{d^3q}{(2\pi)^3} \dfrac{e^{i\bq\cdot(\bx_1-\bx_2)}}{q(q+E_d)}\nonumber\\
  &&\times
   \langle 0^+_F\vert {\cal J}^\dagger_{L,\mu}(\bx_1){\cal J}^{\mu\dagger}_{L}(\bx_2)\vert 0^+_I\rangle,
 \eeqn
 where the nuclear radius $R=1.2A^{1/3}$ is introduced to make the NME dimensionless. $E_d$ is the average energy of intermediate states. Substituting the standard expression ${\cal J}^\dagger_{L,\mu}$ for the one-body charge-changing nuclear current, one finds that the NME is composed of five terms: vector coupling (VV), axial-vector coupling (AA), interference of the axial-vector and induced pseudoscalar coupling (AP), the induced pseudoscalar coupling (PP), and weak-magnetism coupling (MM)  terms, which are related to the products of two current operators ${\cal J}^\dagger_{L,\mu}(\bx_1) {\cal J}^{\mu\dagger}_{L}(\bx_2)$ with the following forms~\cite{Song14},
\begin{subequations}\label{twocurrentR}
\begin{eqnarray}
\label{VV}
VV: &&g_V^2(\bm q^2)\left(\bar\psi\gamma_\mu\tau_-\psi\right)^{(1)}\left(\bar\psi\gamma^\mu\tau_-\psi\right)^{(2)},\\
\label{AA}
AA: &&g_A^2(\bm q^2)\left(\bar\psi\gamma_\mu\gamma_5\tau_-\psi\right)^{(1)}\left(\bar\psi\gamma^\mu\gamma_5\tau_-\psi\right)^{(2)},\\
\label{AP}
AP: &&2g_A(\bm q^2)g_P(\bm q^2)\left(\bar\psi\bm \gamma\gamma_5\tau_-\psi\right)^{(1)}\left(\bar\psi \bm q\gamma_5\tau_-\psi\right)^{(2)},\\
\label{PP}
PP: &&g_P^2(\bm q^2)\left(\bar\psi \bm q\gamma_5\tau_-\psi\right)^{(1)}\left(\bar\psi \bm q\gamma_5\tau_-\psi\right)^{(2)},\\
\label{MM}
MM: &&g_M^2(\bm q^2)\left(\bar\psi\frac{\sigma_{\mu i}}{2m_N}q^i\tau_-\psi\right)^{(1)}\left(\bar\psi\frac{\sigma^{\mu j}}{2m_N}q_j\tau_-\psi\right)^{(2)},
\end{eqnarray}
\end{subequations}
respectively, where $q^\mu$ is the momentum transferred from leptons to nucleons, $\tau_-$ is the isospin lowering operator that changes neutrons into protons, and $\sigma_{\mu\nu}=\frac{\ti}{2}\left[\gamma_\mu,\gamma_\nu\right]$. Following Ref.~\cite{Simkovic1999}, the form factors $g_V(\bm q^2)$, $g_A(\bm q^2)$, $g_M(\bm q^2),$ and $g_P(\bm q^2)$ are chosen as $g_V(\bq^2) =\dfrac{g_V(0)}{(1+\bq^2/\Lambda^2_V)^2}$, $g_A(\bq^2) = \dfrac{g_A(0)}{(1+\bq^2/\Lambda^2_A)^2}$, $g_P(\bq^2) = g_A(\bq^2)\dfrac{2m_N }{\bq^2+m^2_\pi} (1-\dfrac{m^2_\pi}{\Lambda^2_A})$, and $g_M(\bq^2)= (\mu_p-\mu_n) g_V(\bq^2)$,  with $g_V(0) = 1.0$, $g_A(0) = 1.254$,  $\mu_p-\mu_n=3.70$, the cutoff $\Lambda^2_V = 0.710$ (GeV)$^2$, $\Lambda_A = 1.09$ GeV, and the masses $m_N=0.93827$ GeV and $m_\pi=0.13957$ GeV for proton and pion, respectively.

  \section{Numerical details}
  \label{numerical}
The mean-field wave functions $\vert \beta\rangle$ in Eq. (\ref{GCM:wf}) are generated by the RMF calculation based on the point-coupling EDF PC-PK1~\cite{Zhao10}. Pairing correlations between nucleons are treated with the BCS approximation using a density-independent $\delta$ force $V^{pp}_\tau(\br_1,\br_2)=V^{pp}_0\delta(\br_1-\br_2)$ supplemented with an energy-dependent cutoff factor. The pairing strength parameter $V^{pp}_0$ is $-314.550$ MeV fm$^3$ and $-346.500$ MeV fm$^3$ for neutrons and protons, respectively, which were determined by fitting to the neutron and proton average pairing gaps in $^{150}$Nd, $^{150}$Sm provided by the separable finite-range pairing force~\cite{Tian2009}. These paring strength parameters are kept the same for all the $0\nu\beta\beta$ candidate nuclei. More details about the BMF-CDFT calculation can be found in Ref.~\cite{Yao10}.

In the calculation of the NME $M^{0\nu}$, closure approximation is adopted with the average energy of intermediate states given by $E_d=1.12A^{1/2}$~\cite{Haxton84}. 
 All the terms in Eq. (\ref{twocurrentR}) are fully incorporated in the relativistic framework. The finite-nucleon-size (FNS) correction is taken care of by the momentum dependent form factors in Eq. (\ref{twocurrentR}). According to the recent studies based on the unitary correlation operator method  (UCOM)~\cite{Kortelainen07,Menendez2009,Simkovic2009,Engel2009}, the short range correlation (SRC) has a marginal reduction effect ($<10\%$) on the NME for light neutrinos.  To include it would considerably complicate our computational procedure, and therefore we omit the SRC contribution in the present study. More numerical details about the calculation of the NME within the BMF-CDFT have been introduced in Ref.~\cite{Song14}.

\section{Results and discussions}
\label{result}

 \subsection{Structural properties of low-lying states}
\begin{figure}[]
  \centering
 \includegraphics[width=9cm]{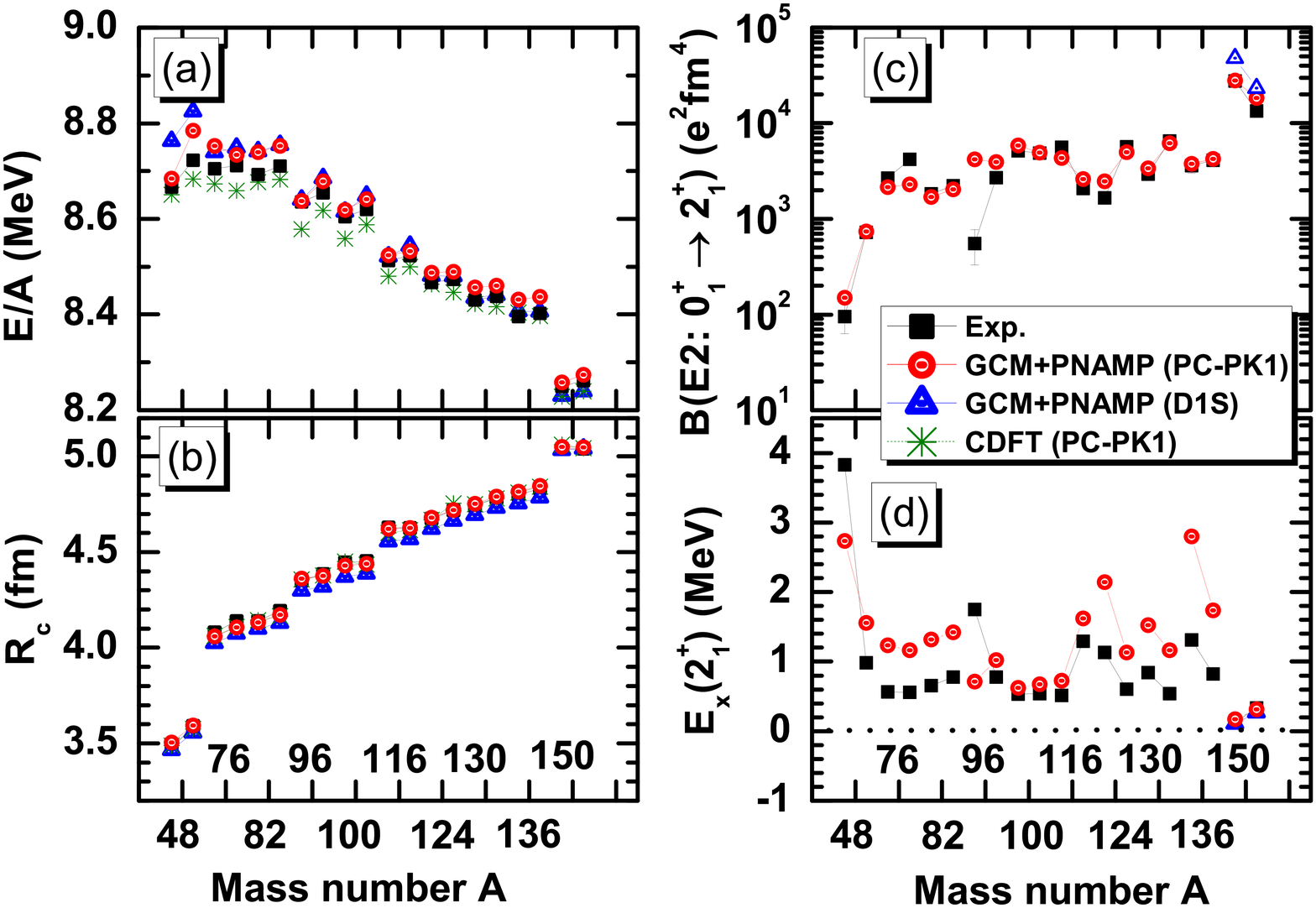}
 \caption{(Color online) Results of the beyond mean-field CDFT (GCM+PNAMP) calculations for the properties
of low-lying states for the $0\nu\beta\beta$ decay candidate nuclei $^{48}$Ca-Ti, $^{76}$Ge-Se, $^{82}$Se-Kr, $^{96}$Zr-Mo, $^{100}$Mo-Ru, $^{116}$Cd-Sn, $^{124}$Sn-Te, $^{130}$Te-Xe, $^{136}$Xe-Ba, and  $^{150}$Nd-Sm, including (a) binding energy and (b) charge radius of correlated $0^+_1$ ground state, as well as (c) $E2$ transition strength $B(E2:0^+_1\to 2^+_1)$ and (d) excitation energy of $2^+_1$ state $E_x(2^+_1)$. The binding energy and charge radius obtained with the pure mean-field CDFT calculation, as well as the available results from the beyond mean-field DFT (GCM+PNAMP) calculation based on the non-relativistic D1S force are given for comparison~\cite{Rodriguez2010}. The experimental data of binding energy and charge radius are taken from Ref.~\cite{Audi03} and Ref.~\cite{Angeli04}, respectively. The data of $B(E2)$ and  $E_x(2^+_1)$ are taken from the NNDC web site~\cite{NNDC}.}
 \label{fig1}
\end{figure}

  We first examine the reliability of our BMF-CDFT calculation for the structural properties of the ten pairs of $0\nu\beta\beta$ decay candidate nuclei $^{48}$Ca-Ti, $^{76}$Ge-Se, $^{82}$Se-Kr, $^{96}$Zr-Mo, $^{100}$Mo-Ru, $^{116}$Cd-Sn, $^{124}$Sn-Te, $^{130}$Te-Xe, $^{136}$Xe-Ba, and  $^{150}$Nd-Sm, which are compared with the results of the BMF-DFT calculations based on the non-relativistic Gogny D1S force~\cite{Rodriguez2010}, and with the data in Fig.~\ref{fig1}. A good agreement with the data is found in both relativistic and non-relativistic BMF calculations for all the candidate nuclei except for the doubly closed-subshell nucleus $^{96}$Zr. For this nucleus, the data of high $E_x(2^+_1)$ and weak $B(E2:0^+_1\to 2^+_1)$ indicate the pronounced proton $Z=40$ and $N=56$ subshells. Actually, the overestimation of the collectivity in $^{96}$Zr is a common problem of most EDF based GCM or collective Hamiltonian calculations~\cite{Delaroche2010,Mei2012}. The lowest excited states in $^{96}$Zr are of particle-hole type~\cite{Molnar89}, the description of which requires the inclusion of noncollective configurations.

\begin{figure}[]
  \centering
 \includegraphics[width=8cm]{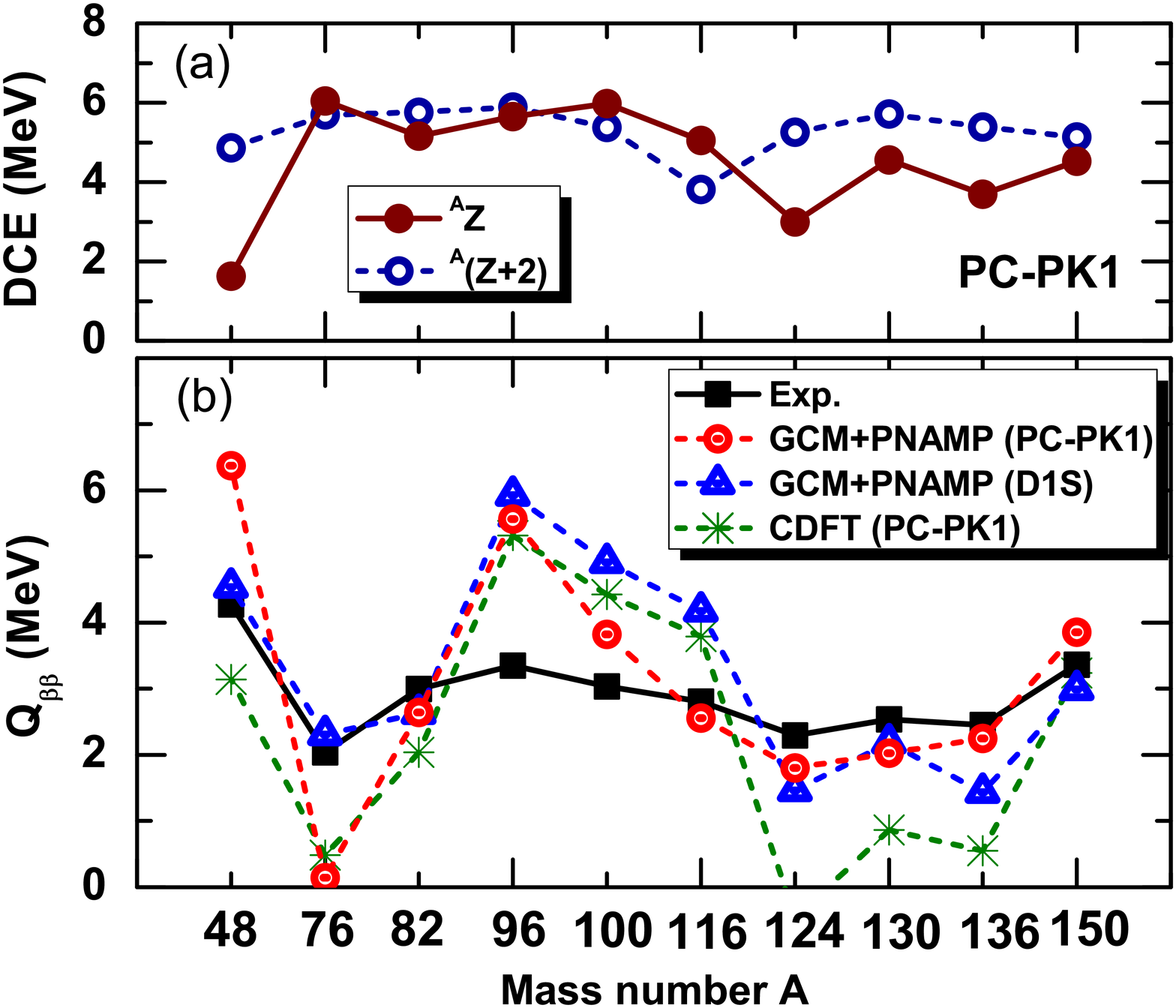}
 \caption{(Color online) (a) Dynamic correlation energy (DCE), $E_{\rm CDFT}-E(0^+)$, and (b) $Q_{\beta\beta}$ values of the $0\nu\beta\beta$ decay obtained with the beyond mean-field CDFT (PC-PK1) calculation, in comparison with the calculated results based on the non-relativistic D1S force and the experimental data~\cite{Audi03}.}
 \label{Qbb}
\end{figure}

\begin{figure}[t]
  \centering
 \includegraphics[width=8.6cm]{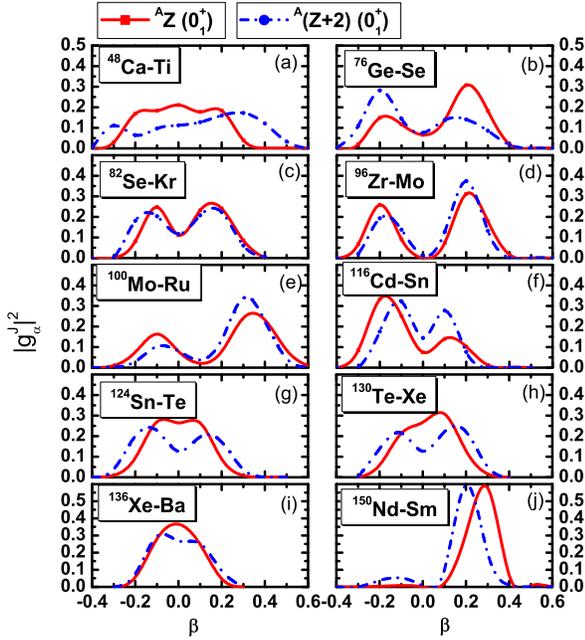}\vspace{-0.5cm}
 \caption{(Color online) Distribution of collective wave functions $\vert g^{J}_\alpha(\beta)\vert^2$ as a function of deformation parameter $\beta$ for the ground state of initial $^AZ$ and final $^A(Z+2)$  nuclei in the $0\nu\beta\beta$ decay.}
 \label{wfs}
\end{figure}

The BMF effects on nuclear binding energy and charge radius can be learnt from Fig.~\ref{fig1}. We define the dynamic correlation energy (DCE) as $E_{\rm CDFT} - E(0^+_1)$, where $E_{\rm CDFT}$ and $E(0^+_1)$ are the total energies of mean-field ground state (global minimum of energy surface) and $0^+_1$ state (correlated ground state), respectively. The DCE ranges from 1.6 MeV to 6.0 MeV, and improves overall the description of binding energies and $Q_{\beta\beta}$ values as depicted in Fig.~\ref{Qbb}. The $Q_{\beta\beta}$ value of $^{48}$Ca has been reproduced in the non-relativistic calculation but is overestimated in our calculation by about 2.0 MeV after taking into account the DCE, which is 1.6 MeV and 4.8 MeV for $^{48}$Ca and $^{48}$Ti, respectively. We note that in our calculation pairing collapse happens around the spherical configuration of $^{48}$Ca, but not in $^{48}$Ti. Therefore, a smaller DCE is gained in $^{48}$Ca than in $^{48}$Ti, which leads to the overestimation of its $Q_{\beta\beta}$. However, pairing collapse is avoided in the non-relativistic calculation where the PNP has also been carried out before variation in the mean-field calculation.
For $^{76}$Ge, on the other hand, the underestimation of the $Q_{\beta\beta}$ might be due to the deficiency of the underlying EDF or the missing of triaxiality, which turns out to be important in the low-lying states~\cite{Bhat14,Sun14}.

Figure~\ref{wfs} displays the distribution of collective wave function $g_\alpha^{J}(\beta)$  for the ground states of initial and final nuclei as a function of the deformation parameter $\beta$, where the $g_\alpha^J(\beta)$ are related to the weight function $f^{JNZ}_\alpha(\beta)$ in Eq. (\ref{GCM:wf}) by the following relation,
\begin{equation}
\label{eq_GCM:30}
g_\alpha^{J}(\beta)
= \sum_{\beta'} \big[ \mathscr{N}^{J}(\beta,\beta') \big]^{1/2} f^{JNZ}_\alpha(\beta'),
\end{equation}
with the overlap kernel, $\mathscr{N}^{J}(\beta,\beta')=\langle  \beta \vert \hat P^J_{00}  \hat P^N\hat P^Z\vert\beta'\rangle$. Since $g_\alpha^{J}(\beta)$s are orthonormal, they can reflect the dominant configuration of each state.
It is seen that the deformations of dominant configurations in the ground state of mother and daughter nuclei  are somewhat different, as already discussed in Refs.~\cite{Rodriguez2010,Song14}. This static deformation effect can quench the NME of $0\nu\beta\beta$ decay significantly in particular for the case where the deformations of the mother and daughter nuclei differ considerably from each other, such as $^{76}$Ge-Se and  $^{150}$Nd-Sm. Moreover, shape fluctuation is shown to be significant in the light $0\nu\beta\beta$ candidate nuclei, the description of which is impossible with the approaches based on single-reference state~\cite{Rath2010,Fang2010,Mustonen2013}. This dynamic deformation effect (or shape mixing effect) could moderate the quenching effect from the static deformation on the NMEs~\cite{Song14}, which is fully taken into account in the present multi-reference BMF-CDFT approach.

 \subsection{Nuclear matrix elements for the $0\nu\beta\beta$ decay}

 \begin{table}[b]
 \centering
 \tabcolsep=4pt
 \caption{The normalized NME $\tilde M^{0\nu}$ for the $0\nu\beta\beta$-decay obtained with the particle number projected spherical mean-field configuration ($\beta_I=\beta_F=0$) by the PC-PK1 force using both the relativistic and non-relativistic reduced (first-order of $q/m_p$ in the one-body current) transition operators.  The ratio of the $AA$ term to the total NME, $R_{AA}\equiv \tilde M^{0\nu}_{AA}/\tilde M^{0\nu}$, the relativistic effect $\Delta_{\rm Rel.}\equiv (\tilde M^{0\nu}-\tilde M^{0\nu}_{\rm NR})/\tilde M^{0\nu}$ and the ratio of the tensor part  to the total NME, $R_T\equiv \tilde M^{0\nu}_{\rm NR, T}/\tilde M^{0\nu}_{\rm NR}$, are also presented.}
 \begin{tabular}{lrrrrr}
  \hline\hline
  {\rm  Sph+PNP (PC-PK1)} &    $\tilde M^{0\nu}$ & $R_{AA}$ &  $\tilde M^{0\nu}_{\rm NR}$  & $\Delta_{\rm Rel.}$ & $R_T$  \\
   \hline
 $^{48}$Ca $\to ^{48}$Ti &     3.66  & 81\%  &	3.74 & $-$2.1\%&  $-$2.4\%   \\
 $^{76}$Ge $\to ^{76}$Se &     7.59  & 94\%  & 7.71  &$-$1.6\% &   3.5\%     \\
 $^{82}$Se $\to ^{82}$Kr &     7.58  & 93\%  & 7.68  & $-$1.4\%&  2.9\%    \\
 $^{96}$Zr $\to ^{96}$Mo &     5.64	 & 95\%  &  5.63 &0.2\%    &  3.6\%    \\
 $^{100}$Mo $\to ^{100}$Ru &   10.92 & 95\%  & 10.91 &0.1\%    &   3.5\%  \\
 $^{116}$Cd $\to ^{116}$Sn &   6.18  & 94\%  & 6.13  &0.7\%    &  1.9\%   \\
 $^{124}$Sn $\to ^{124}$Te &   6.66  & 94\%  & 6.78  &$-$1.8\% &  4.9\%   \\
 $^{130}$Te $\to ^{130}$Xe &    9.50  & 94\% & 9.64  &$-$1.4\% &  4.3\%   \\
 $^{136}$Xe $\to ^{136}$Ba &    6.59  & 94\% & 6.70  &$-$1.7\% & 4.1\%      \\
 $^{150}$Nd $\to ^{150}$Sm &    13.25 & 95\% & 13.08 & 1.3\%   &  2.5\%       \\
 \hline \hline
\end{tabular}
 \label{tab1}
\end{table}

 In order to show the deformation-dependence of the NME, Table~\ref{tab1} presents the normalized NME $\tilde{M}^{0\nu}(\beta_I, \beta_F)$ at spherical shape ($\beta_I=\beta_F=0$) for the $0\nu\beta\beta$-decay obtained with both the relativistic and non-relativistic reduced transition operators, where $\tilde M^{0\nu}$ is defined as
\begin{equation}\label{MIFnorm}
\tilde{M}^{0\nu}(\beta_I, \beta_F) = {\mathcal N}_F{\mathcal N}_I\,\langle \beta_F|\hat{\mathcal O}^{0\nu}\hat P^{J=0} \hat P^{N_I}\hat P^{Z_I}| \beta_I\rangle,
\end{equation}
with ${\mathcal N}^{-2}_a= \langle \beta_a|\hat  P^{J=0}_{00}\hat P^{N_a}\hat P^{Z_a}| \beta_a\rangle$ for $a=I,F$.
It is seen that the error arisen from the first-order non-relativistic reduction is marginal, which can either increase or decrease the total NME by a factor within $2\%$. This value is modified only slightly in the full GCM calculation, for instance becoming $\sim5\%$ for $^{150}$Nd~\cite{Song14}. The one-body charge-changing nucleon current, Eq.~(\ref{twocurrentR}), generates not only the Fermi and Gamow-Teller (GT) terms but also tensor terms that have been neglected in the non-relativistic study~\cite{Rodriguez2010}. With the help of non-relativistic approximation of the transition operator, one can isolate the contribution of the tensor part~\cite{Simkovic1999,Song14}, which is obtained by subtracting the contributions of Fermi and GT terms from the total NME. It is shown in Table~\ref{tab1} that the contribution of tensor terms is within $5\%$ of the total NME.

\begin{figure}[hbt]
  \centering
 \includegraphics[width=9cm]{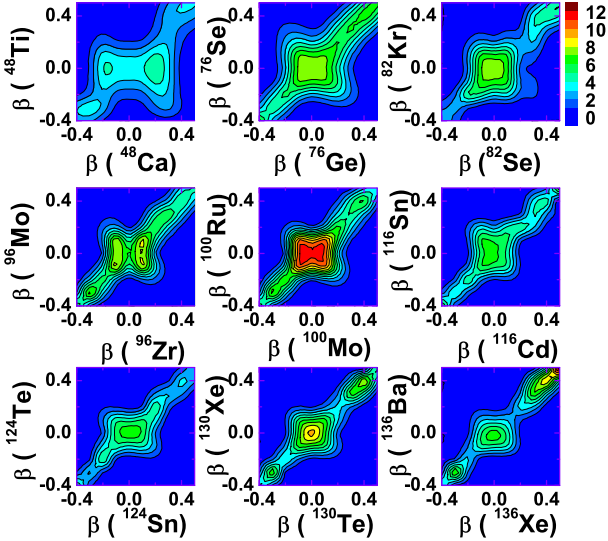}
 \caption{(Color online) Normalized NME $\tilde M^{0\nu}$ as a function of the intrinsic deformation parameter $\beta$ of the initial $^AZ$ and final $^A(Z+2)$  nuclei.}
 \label{NME:normalized}
\end{figure}

Figure~\ref{NME:normalized} displays the normalized NME $\tilde M^{0\nu}$ as a function of the intrinsic quadrupole deformation $\beta_I$ and $\beta_F$ of the mother and daughter nuclei, respectively. Similar to the behavior of the GT part shown in the MR-DFT (D1S) calculation~\cite{Rodriguez2010}, the normalized NME $\tilde M^{0\nu}$ is concentrated rather symmetrically along the diagonal line $\beta_I=\beta_F$, implying that the decay between nuclei with different deformation is strongly hindered. Moreover, the $\tilde M^{0\nu}$ has the largest value at the spherical configuration for most candidate nuclei except for $^{48}$Ca-Ti, $^{96}$Zr-Mo, and $^{136}$Xe-Ba. It implies that generally the $0\nu\beta\beta$-decay is favored if both nuclei are spherical.  The largest $\tilde M^{0\nu}$ in $^{136}$Xe-Ba is found around the deformation region with $\beta_I=\beta_F\simeq0.5$, at which deformed configuration, pairing energy is peaked in both nuclei due to the very high single-particle level density. However, this configuration ($\beta\simeq0.5$) has a negligible contribution to the final NME of $^{136}$Xe-Ba because its weight is almost zero in the ground-state wave function, cf.~Fig.~\ref{wfs}.

\begin{figure}[]
  \centering
 \includegraphics[width=9cm]{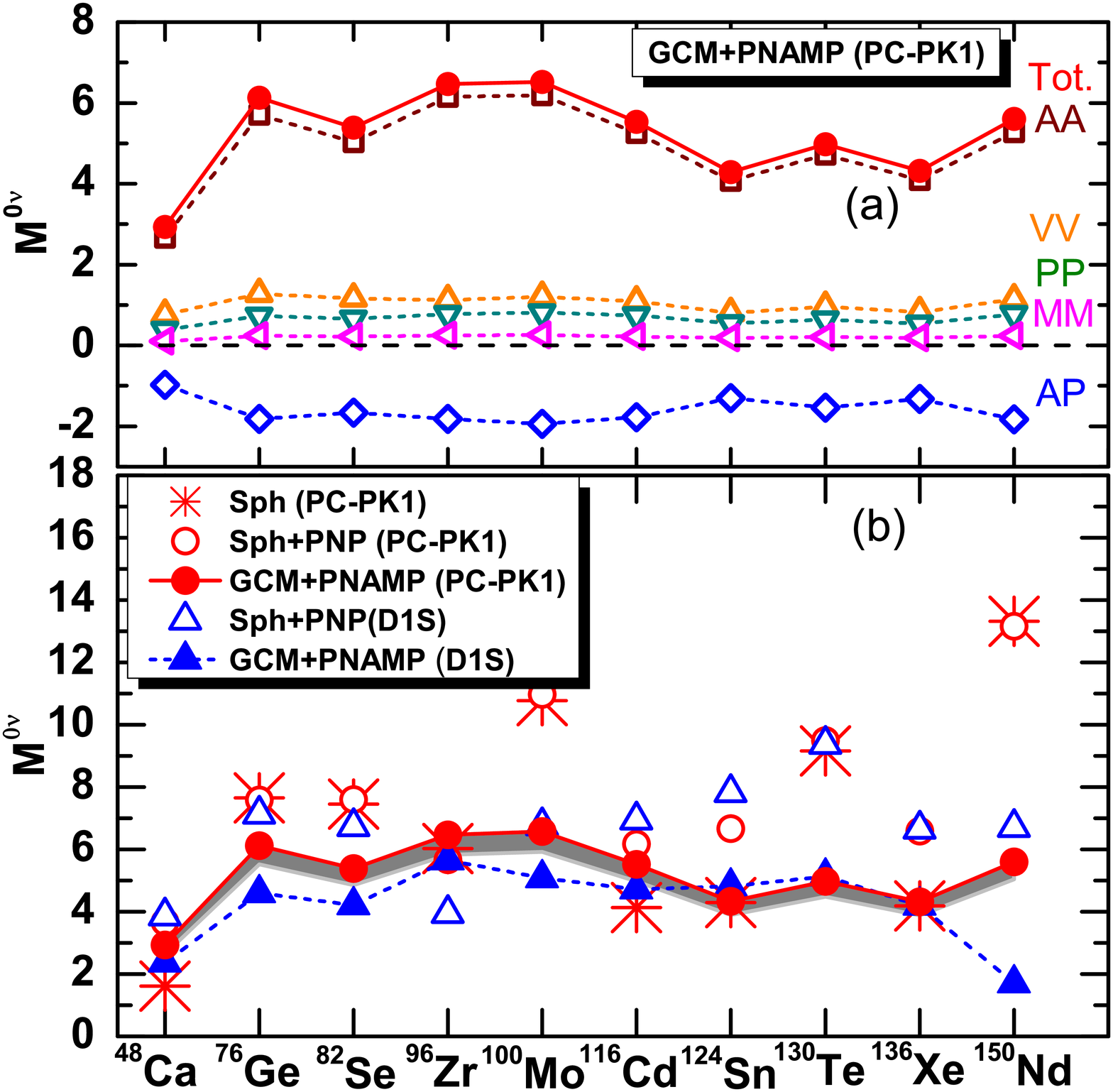}
 \caption{(Color online) (a) Decomposition of the total NMEs from the final GCM+PNAMP (PC-PK1) calculation; (b) the total NMEs calculated with either only spherical configuration or full configurations, in comparison with those of GCM+PNAMP (D1S) from Ref.~\cite{Rodriguez2010}. The shaded area indicates the uncertainty of the SRC effect within $10\%$. See text for more details.}
 \label{NME:GCM}
\end{figure}

\begin{figure}[]
  \centering
 \includegraphics[width=9cm]{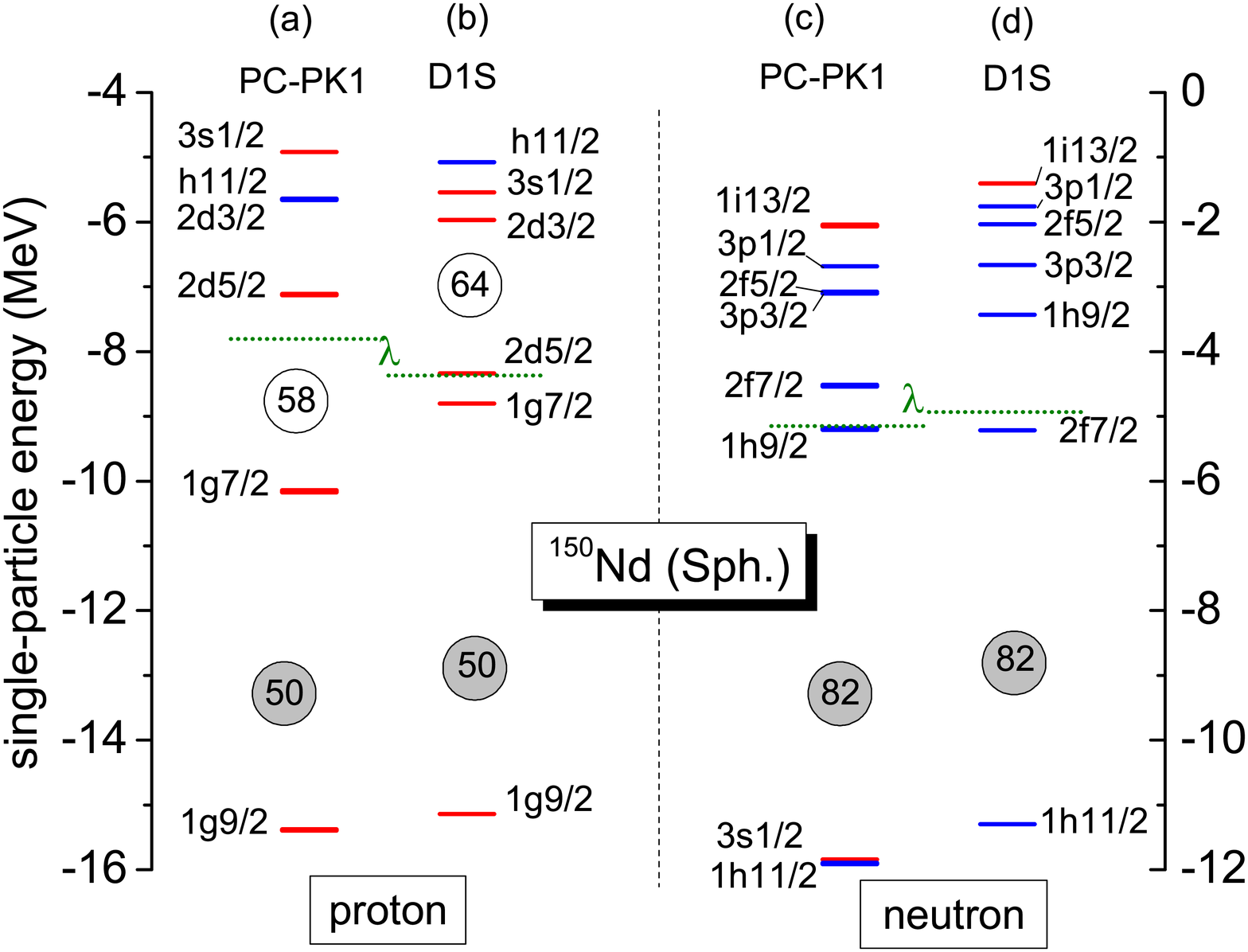}
 \caption{(Color online) The single-particle energy levels of neutrons and protons in $^{150}$Nd predicted by the PC-PK1 force, in comparison with that by the Gogny D1S force taken from Ref.~\cite{CEA}. }
 \label{spe:comparison}
\end{figure}

\begin{figure}[]
  \centering
 \includegraphics[width=9cm]{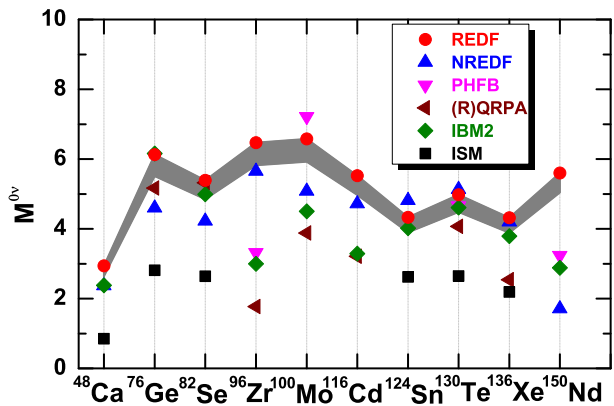}
 \caption{(Color online) Comparison of the NME $M^{0\nu}$ for the $0\nu\beta\beta$-decay from different model calculations. The shaded area indicates the uncertainty of the SRC effect within $10\%$. }
 \label{NME:comparison}
\end{figure}

 Figure~\ref{NME:GCM}(a) displays the contribution of each coupling term ($AA, VV, PP, MM, AP$) in Eq.(\ref{twocurrentR}) to the total NMEs. It is shown that the weak-magnetism ($MM$) term is negligible ($\sim4\%$). The interference term  ($AP$) of the axial-vector and pseudoscalar coupling has an opposite contribution ($\sim30\%$), which almost cancels out the sum of $VV$, $PP$, and $MM$ terms. Of particular interest is that the total NME has a very similar behavior as that of the predominated $AA$ term with the ratio $R_{AA}\simeq95\%$. Actually, we have found that the deformation-dependent NMEs shown in Fig.~\ref{NME:normalized} are also very similar even if we include only the $AA$ term. It indicates that the $AA$ term provides a good approximation for the total NME, Eq.(\ref{NME:formula}). In the non-relativistic approximation, the two-current operator with only the axial-vector coupling term is simplified as ${\cal J}^\dagger_{L,\mu}(\bx_1) {\cal J}^{\mu\dagger}_{L}(\bx_2)=-g^2_A(q^2) \bm\sigma^{(1)}\cdot\bm\sigma^{(2)}\tau_-^{(1)}\tau_-^{(2)}$, the calculation of which is much cheaper than computing the full terms, cf.~(\ref{twocurrentR}). Similar conclusion can also be made based on the results of QRPA calculation~\cite{Simkovic1999} using the non-relativistic reduced operators. Figure~\ref{NME:GCM}(b) displays the NMEs calculated either with pure spherical configuration or with full configurations in the GCM+PNAMP (PC-PK1), in comparison with those of  the non-relativistic results~\cite{Rodriguez2010}. Before comparing the two results, we should point out that in the non-relativistic calculation~\cite{Rodriguez2010}, the SRC effect was taken into account with the UCOM, while the tensor terms were neglected. These two effects can bring a difference up to $\sim15\%$ in the NMEs. By taking into account this point, one can draw the conclusion from Fig.~\ref{NME:GCM}(b) that these two calculations give consistent results for  the total NMEs for all the candidate nuclei with the exception of $^{150}$Nd.

 As we have already discussed in Ref.~\cite{Song14}, one of the reasons leading to the large discrepancy in the NMEs of $^{150}$Nd is the different distribution of the collective wave function for the ground states of $^{150}$Nd and $^{150}$Sm. Besides, the normalized NMEs at the spherical configuration differ from each other by a factor of about two, which might be due to the different level density around Fermi surface, giving rise to very different pairing properties. Figure~\ref{spe:comparison} displays a comparison of the  single-particle energy levels of neutrons and protons at the spherical configuration of $^{150}$Nd obtained by the PC-PK1 and Gogny D1S forces. In contrast to results by the D1S force~\cite{CEA,Rodriguez2012} but similarly to other CDFT calculations~\cite{Long2009}, the PC-PK1 force predicts a large $Z=58$ shell gap and a somewhat small $Z=64$ gap, which is however not supported by the experiment~\cite{Nagai81}. Notice that the large $Z=64$ shell gap can be reproduced by including the non-local exchange terms of the isoscalar field couplings in the relativistic Hartree-Fock (RHF) calculation, which is able to give an enhanced spin-orbit splitting for the proton $2d$ states~\cite{Long2009}.   It remains an open question whether the NME for $^{150}$Nd is reduced  or not by including these terms in the RHF calculations. In short, the large discrepancy in the NMEs of $^{150}$Nd by the PC-PK1 and D1S forces is the result of the interplay of pairing correlations and the underlying shell structure. A comprehensive understanding of this discrepancy requires further dedicated investigations.

 \begin{table*}[t]
 \centering
 \tabcolsep=8pt
 \caption{The calculated NME $M^{0\nu}$ of the $0\nu\beta\beta$-decay with the REDF (PC-PK1), in comparison with those by the NREDF (D1S), RQRPA, PHFB, ISM,  and IBM2. Only the results considering the short-range correlation (SRC) effect by UCOM, except for the IBM2 where CCM is used and using the parameter $R = 1.2 A^{1/3}$ fm are adopted for comparison. The values in the parenthese are the results with additional pairing fluctuations.}
 \begin{tabular}{ccccccc}
  \hline\hline
 Models &  REDF(PC-PK1) & NREDF(D1S) & RQRPA (T\"{u}bingen) & PHFB & ISM & IBM2 \\
  $g_A(0)$ &  1.254  & 1.25  & 1.254   &  1.254  &  1.25  & 1.269 \\
 \hline
 $^{48}$Ca  $\to ^{48}$Ti  & 2.94   & 2.37 (2.23)   &                    &         &  0.85   & 2.38  \\
 $^{76}$Ge  $\to ^{76}$Se  & 6.13   & 4.60 (5.55)   &        5.17        &         &  2.81   & 6.16  \\
 $^{82}$Se  $\to ^{82}$Kr  & 5.40   & 4.22 (4.67)  &        5.32        &         &  2.64   &  4.99  \\
 $^{96}$Zr  $\to ^{96}$Mo  & 6.47   & 5.65 (6.50)  &        1.77        &  3.32   &         &  3.00   \\
 $^{100}$Mo $\to ^{100}$Ru & 6.58   & 5.08  (6.59) &        3.88        &  7.22   &         &  4.50  \\
 $^{116}$Cd $\to ^{116}$Sn & 5.52   & 4.72  (5.35) &        3.21        &         &         &  3.29  \\
 $^{124}$Sn $\to ^{124}$Te & 4.33   & 4.81  (5.79) &                    &         &  2.62   &  4.02    \\
 $^{130}$Te $\to ^{130}$Xe &4.98    & 5.13  (6.40) &        4.07        &  4.66   &   2.65  &  4.61  \\
 $^{136}$Xe $\to ^{136}$Ba &4.32    & 4.20  (4.77) &        2.54        &         &   2.19  &  3.79  \\
 $^{150}$Nd $\to ^{150}$Sm & 5.60   & 1.71  (2.19) &                    &  3.24   &         &  2.88 \\
 \hline \hline
\end{tabular}
 \label{tab2}
\end{table*}

  The effect of PNP on the NMEs for the $0\nu\beta\beta$-decay is also shown in Fig.~\ref{NME:GCM}(b). In the calculation with pure spherical configuration, the PNP increases significantly the NMEs  evolved with one (semi)magic nucleus, including $^{48}$Ca ($127\%$), $^{116}$Cd ($49\%$), $^{124}$Sn ($55\%$), and $^{136}$Xe ($58\%$), where pairing collapse occurs in either protons or neutrons. The increase in the NMEs by the PNP is mainly through the superfluid partner nucleus. For $^{48}$Ca, pairing collapse is found in both neutrons and protons, leading to about twice enhanced normalized NME than the other three ones. It can be understood from Eq.(\ref{MIFnorm}) that the $\langle \beta_F=0|\hat{\mathcal O}^{0\nu}\hat P^{J=0} \hat P^{N_I}\hat P^{Z_I}| \beta_I=0\rangle$ for $^{48}$Ca-Ti does not change by the PNP, while the normalization factor ${\mathcal N}_F$ for the daughter nucleus $^{48}$Ti is increased, resulting in the enhanced normalized NME.


Figure~\ref{NME:comparison} displays our final NMEs for the  $0\nu\beta\beta$-decay in comparison with those by the ISM~\cite{Menendez2009}, renormalized QRPA (RQRPA)~\cite{Faessler12}, PHFB~\cite{Rath2010}, NREDF (D1S)~\cite{Rodriguez2010}, and the IBM2~\cite{Barea2009}. There are also other calculations that are not taken for comparison. Here, only the calculations considering the SRC effect with the UCOM (except for the IBM2 calculation with the
coupled-cluster model (CCM)) and using the radius parameter $R = 1.2 A^{1/3}$ fm are adopted for comparison. The values are given in Tab.~\ref{tab2}. Our results are amongst the largest values of the existing calculations in most cases, except for $^{100}$Mo-Ru, $^{124}$Sn-Te and $^{130}$Te-Xe. Moreover, the NME for $^{96}$Zr in both EDF-based calculations is significantly larger than the other results, which can be traced back to the overestimated collectivity. If the ground state of $^{96}$Zr was taken as the pure spherical configuration, the NME becomes 5.64 (PC-PK1) and 3.94 (D1S), respectively. We note that the consideration of higher-order deformation in nuclear wave functions, such as octupole deformation in $^{150}$Sm-Nd~\cite{Zhang10,Bvumbi13}, and triaxiality in $^{76}$Ge-Se~\cite{Bhat14,Sun14} and $^{100}$Mo-Ru~\cite{Wrzosek-Lipska12}, is expected to hinder the corresponding NMEs further in the DFT calculation.


 \begin{table}[tb]
\centering
\tabcolsep=2pt
 \caption{The upper limits of the effective neutrino mass $\langle m_{\beta\beta}\rangle$ (eV) based on the NMEs from the present GCM+PNAMP (PC-PK1) calculation, the lower limits of the half-life $T^{0\nu}_{1/2}(\times 10^{24}$ yr) for the $0\nu\beta\beta$-decay from most recent measurements~\cite{Umehara08,Agostini13,Barabash11,Andreotti11,Auger12,Gando13,Argyriades09} and the phase-space factor $G_{0\nu} (\times 10^{-15}$ yr$^{-1}$) from Ref.~\cite{Kotila12}.}
 \begin{tabular}{ccccccccc}
  \hline\hline
                 & $^{48}$Ca  & $^{76}$Ge & $^{82}$Se &    $^{100}$Mo & $^{130}$Te  & $^{136}$Xe &  $^{150}$Nd \\
 \hline
   $\langle m_{\beta\beta}\rangle$
   &  $\leq$ 2.92 & $\leq$ 0.20 &  $\leq$  1.00 & $\leq$  0.38  & $\leq$ 0.33 & $\leq$ 0.11 & $\leq$ 1.72 \\
  $T^{0\nu}_{1/2}$ & $\geq$ 0.058 & $\geq$ 30 & $\geq$ 0.36 & $\geq$ 1.1 & $\geq$ 2.8  & $\geq$ 34 & $\geq$ 0.018 \\
  $G_{0\nu}$       & 24.81 & 2.363 & 10.16  & 15.92 & 14.22  &14.58& 63.03\\
 \hline \hline
\end{tabular}
 \label{tab3}
\end{table}

Table~\ref{tab3} lists the upper limits of the effective neutrino mass $\langle m_{\beta\beta}\rangle$ based on the present calculated NMEs for the nuclei whose lower limits of the half-life $T^{0\nu}_{1/2}$ for the $0\nu\beta\beta$-decay have been recently measured~\cite{Umehara08,Agostini13,Barabash11,Andreotti11,Gando13,Argyriades09}. The smallest value ($\leq0.11$ eV) for the upper limit $\langle m_{\beta\beta}\rangle$  is found  based on the combined results from KamLAND-Zen~\cite{Gando13} and EXO-200~\cite{Auger12} collaborations for the$0\nu\beta\beta$-decay half-life ($T^{0\nu}_{1/2}\geq3.4\times10^{25}$ yr at 90\% confidence level) of $^{136}$Xe. This value is closest to but still larger than the estimated value ($20-50$ meV  based on the inverted hierarchy for neutrino masses~\cite{Bilenky12}) by a factor of $2-5$.

\section{Summary and outlook}
\label{Summary}

We have reported a systematic study of NMEs for the $0\nu\beta\beta$-decay candidate nuclei with our state-of-the-art BMF-CDFT, where the NMEs have been calculated with the full relativistic transition operators derived from one-body charge-changing nuclear current. The effects of PNP+AMP as well as the static and dynamic deformations in the nuclear wave functions have been taken into account automatically. The reliability of the nuclear wave functions has been examined by comparing the calculated low-energy structural properties with the corresponding data.

The novel findings in the present systematic study are summarized as follows:
\begin{itemize}
  \item In most of the cases there is a much better agreement between NREDF and REDF calculated matrix elements than the case of $^{150}$Nd~\cite{Song14}, which requires further dedicated investigations. It indicates that in general the NMEs are not much sensitive to the underlying EDF.
  \item The axial-vector coupling ($AA$) term exhausts more than 95\% of the total NME, which provides an economical way to calculate the total NME in the future.
  \item The net contribution from relativistic effect and tensor terms that have been neglected in the NREDF study of Ref.~\cite{Rodriguez2010} turns out to be within $10\%$ for all the candidate nuclei. The PNP effect that was neglected in most QRPA and PHFB studies increases significantly the NME for the $0\nu\beta\beta$-decay where magic or semi-magic nuclei are evolved. The net effects of static and dynamic deformation turn out to reduce significantly the NME for most candidate nuclei.
  \item The smallest upper limit on $\langle m_{\beta\beta}\rangle$($\leq0.11$ eV) has been found based on the latest data on the $0\nu\beta\beta$-decay of $^{136}$Xe.
\end{itemize}
Finally, we point out that the effect of higher-order deformation needs to be studied in this framework. Moreover, the quenching effect of two-body currents~\cite{Menendez2011,Engel2014} and enhancement effect of pairing fluctuation~\cite{Vaquero2013} on the NME are comparable in size, that is, $10\%-40\%$. These two effects may not cancel out exactly. The fluctuation effect in proton-neutron pairing amplitude also has influence on the value of NME~\cite{Hinohara14}. Therefore, a more careful study within the BMF-CDFT by taking into account all these effects in a unified way needs to be carried out in the near future.

%
\section*{Acknowledgements} This work was supported by the Tohoku University Focused Research Project ``Understanding the origins for matters in universe", the Major State 973 Program 2013CB834400, the NSFC under Grant Nos. 11175002, 11105111, 11335002, and 11305134, the Fundamental Research Fund for the Central Universities (XDJK2013C028), the Overseas Distinguished Professor Project from Ministry of Education (MS2010BJDX001), and the DFG Cluster of Excellence \textquotedblleft Origin and Structure of the Universe\textquotedblright\ (www.universe-cluster.de).


\end{document}